\documentclass[a4paper,12pt,english,german]{article}

\usepackage{caption}
\usepackage{subcaption}

\usepackage{mathtools}
\usepackage{graphicx}

\usepackage{amssymb,amsmath,scalefnt,cite}
\usepackage{titlesec}

\begin{document}

\centerline{\bf Quantum Brownian oscillator for the stock market}

\bigskip

\centerline{Jasmina Jekni\' c-Dugi\' c$^a$, Sonja Radi\' c$^b$, Igor Petrovi\' c$^c$, Momir Arsenijevi\' c$^b$, Miroljub Dugi\' c$^{\ast b}$}

\bigskip

$^a$University of Ni\v s, Faculty of Science and Mathematics, Vi\v
segradska 33, 18000 Ni\v s, Serbia

$^b$University of Kragujevac, Faculty of Science, Radoja
Domanovi\' ca 12, 34000 Kragujevac, Serbia

$^c$Svetog Save 98, 18230 Sokobanja, Serbia

\bigskip

{\bf Abstract}

\smallskip

\noindent We pursue the quantum-mechanical challenge to the efficient market hypothesis for the stock market by employing the quantum Brownian motion model.
We utilize the quantum Caldeira-Leggett master equation as a possible phenomenological model for the stock-market-prices fluctuations while introducing the external harmonic field for the
Brownian particle. Two quantum regimes are of particular interest: the exact regime as well as the approximate regime of the pure decoherence ("recoilless") limit of the Caldeira-Leggett equation.
By calculating the standard deviation and the  kurtosis for the particle's position observable, we can detect deviations of the quantum-mechanical behavior from the
classical counterpart, which bases the efficient market hypothesis.
By varying the damping factor, temperature as well as the oscillator's frequency, we are able to provide interpretation of different economic scenarios and possible situations
that are not normally recognized by the efficient market hypothesis. Hence we recognize the quantum Brownian oscillator as a possibly useful model for the realistic behavior of stock prices.

\bigskip


\bigskip

$^{\ast}$Corresponding author: mdugic18@sbb.rs

\bigskip

{\bf 1. Introduction}

\bigskip

Deviation from the Gaussian (so-called normal) distribution of returns appears as a universal {\it empirical} fact for  different
markets, ranging from the markets in developed countries, such as Germany [1] and US [2, 3], to markets in developing countries, like India [4] and China [5].
It manifests as, inter alia,
the "fat tails" as well as the positive excess kurtosis for the probability distribution of returns. The distribution is expected to converge to the
standard Gaussian behavior after sufficiently long time interval [6-8] thus suggesting the presence of some, hopefully universal, regularities
governing the complex financial systems [9-11].

Appearance of both the fat tail deviation and the positive excess kurtosis exhibit non-Markovian behavior thus making obvious that the stock market
does not satisfy the classical Brownian motion model, which is characteristic of the remarkable efficient market hypothesis (EMH) [12]. As a natural
step forward appear the attempts to properly include certain quantum  models [13-15] that led to a number of different approaches and models pursuing
the quantum paradigm in the quantitative finance field [15-23]. As a rationale for the quantum mechanical approach it is often invoked the market irrationality--in
sharp contrast to EMH.
Behavioral economists pledge for the important role of the agents irrationality in the realistic stock transactions so much that irrationality might substantially
contribute to the (empirically well-known) persistent fluctuations of the stock price even when there is no information to impel the stock price.
As a possible model of irrationality appears the quantum mechanical uncertainty that is therefore interpreted as the market uncertainty, that drives volatility
[18,20,22] (and the references therein).

Different quantum models have been used, such as the particle in the potential well [19,20], the quantum damped oscillator suggested in [20], harmonic oscillator [21,23], quantum Brownian motion [22] etc.
Quantum Brownian motion is particularly interesting bearing in mind it could be regarded a quantum-mechanical counterpart of the classical Brownian motion, which, in turn, backs the profound
efficient market hypothesis [12]. Hence an elaborated quantum-mechanical challenge, notably Ref. [22], for the efficient market hypothesis that is worth  pursuit.

In this paper we utilize the quantum Brownian motion as modelled by the well-known Caldeira-Leggett (CL) master equation [22,24,25]. We go beyond the existing models in that we
introduce the external harmonic field for the Brownian particle, while, on the other hand, we pay special attention to the pure decoherence (the so-called "recoilless") limit of the CL master equation. The decoherence limit
regards the off-diagonal terms, $\rho(x,x')=\langle x\vert\hat\rho\vert x'\rangle$, that often remain an open issue [22].
As distinct from the similar approaches, we regard the CL  equation as a "{\it phenomenological}" master equation, meaning that we are not concerned with the underlying microscopic physical details that lead [24,25] to the equation. Rather we investigate usefulness of the CL equation for a harmonic oscillator in the context of the econophysics studies.

Generally, appearance of the external field is an attempt to model the external macroscopic influence on the stock market, such as
the daily price limitation of the stock markets in China [21,22,26] or to distinguish between the "regular" and "irregular" returns in the United Kingdom's Financial Times-Stock Exchange (FTSE) All Share Index [19].
In this regard, we consider our model of the harmonic oscillator Brownian particle as possibly more realistic than the model of the free Brownin motion [22].

By following the standard wisdom, we calculate  the standard deviation and the  kurtosis for both the exact and the recoilless limit of the quantum harmonic Brownian particle and compare the obtained results with the classical counterpart. Comparison regards not only the  damping rate $\gamma$ or  the bath's temperature, but also  the oscillator's frequency $\omega$. Justification of our model of the  harmonic Brownian particle comes also from the similar considerations [21,23] as well as from the analogous model of the (classical) damped harmonic oscillator [27] that was considered as a possible physical basis for the EMH [12]. To the extent that the classical harmonic Brownian particle properly bases the EMH,
we provide some evidences against market efficiency as well as possibly a useful physical (quantum-mechanical) model for the stock market.

In Section 2, we introduce and briefly discuss the model. In Section 3 we provide the main results of this paper.  In Section 4 we provide discussion and conclusion for the obtained results by paying special attention to the
interpretation of our findings in the context of the possible economic scenarios and situations.

\bigskip
{\bf 2. The Caldeira-Leggett master equation for the harmonic oscillator}

\bigskip

The Caldeira-Leggett master equation for quantum state ("density matrix"), $\hat\rho$, of one-dimensional system, in the Schr\" odinger picture, reads [24,25]:

\begin{equation}
{d \hat\rho(t)\over dt} = -{\imath\over\hbar} [\hat H, \hat\rho(t)] -
{\imath\gamma\over\hbar} [\hat x, \{\hat p, \hat\rho(t)\}] - {2m\gamma
k_B T\over \hbar^2}[\hat x, [\hat x, \hat\rho_R(t)]].
\end{equation}

In eq.(1), the only degree of freedom of the particle is the Descartes coordinate $\hat x$, while $\hat p$ stands for its conjugate momentum; the commutation relation, $[\hat x,\hat p]=\hat x\hat p - \hat p\hat x=\imath\hbar$.
The system's Hamiltonian $\hat H$ generates the unitary dynamics described by the first commutator on the rhs of eq.(1), while the second and the third terms
model the quantum mechanical dissipation and decoherence (sometimes also referred to as "dephasing"), respectively, both determined by the non-negative and time-independent
damping coefficient $\gamma$. By $m$ we denote the system's mass, while $k_B$ and $T$ stand for the Boltzmann constant and the  thermal-bath's temperature, respectively.
The system's Hamiltonian (while neglecting the Lamb-shift term)

\begin{equation}
\hat H=\hat T+\hat V = {\hat p^2\over 2m} + V(\hat x),
\end{equation}

\noindent where the external potential $\hat V=0$ describes the free Brownian particle, while $\hat V=m\omega^2\hat x^2/2$ regards the particle in the external harmonic potential with the frequency $\omega$ and the zero
equilibrium position. The square brackets stand for the commutator, while the curly brackets for the anticommutator, $\{A,B\}=AB+BA$.

In this paper we take the equation (1) as given, stipulated, without resorting to the details of its microscopic origin [28].
This gives us a freedom to vary the values of the parameters $\gamma, m, T, \omega$--those variations are limited by the assumptions of large temperature and weak interaction in the microscopic derivation of the CL equation (1) [24,25].

For very large $T$ and/or very large mass $m$, the exact equation (1) can be approximated thus giving rise to the decoherence-limit (the so-called "recoilless limit") [25]. In this limit, the third--the decoherence--term dominates the system's dynamics thus allowing for neglecting the second (the dissipation) term of eq.(1). Then the particle undergoes the environment-induced decoherence [25,29] without dissipation. Bearing in mind that (in the Heisenberg picture) the CL equation (1) has a well-defined classical counterpart in the form of the Langevin equation  [24,25], it is particularly interesting to compare the   decoherence-limit results with the known and exact classical expressions.

Prescription of the general model assumptions into the econophysics context is standard, e.g., [22]: the degree of freedom $\hat x$ regards the (logarithmic) price, the momentum $\hat p$ the trend of the price, while
$m$ now stands for the stock  inertia quantifying the market capitalization. The bath's temperature $T$ quantifies the externally-induced fluctuations while the damping coefficient $\gamma$
quantifies the externally-induced damping strength [22]. Therefore the parameters variations may regard the different scenarios for the stock transactions with respect e.g. to the market capitalization ($m$) and the frequency of the external interventions ($\omega$).

The moments characterizing the distribution $\rho(x,x')$ are all of the form $tr (\hat A\hat\rho(t))$ (with the time-independent $\hat A$ in the Schr\" odinger picture). With the use of the identities, $tr(A[B,C])=tr([A,B]C)$ and $tr(A\{B,C\})=tr(\{A,B\}C)$, it easily follow the differential equations for the moments, in the following general form:

\begin{equation}
{d(tr\hat A\hat\rho(t))\over dt} = {-\imath\over\hbar} tr([\hat A,\hat H]\hat\rho) - {\imath\gamma\over\hbar} tr(\{[\hat A,\hat x],\hat p\}\rho) - {2m\gamma
k_B T\over \hbar^2}tr([\hat x, [\hat x, \hat A]]\hat \rho).
\end{equation}

\bigskip

{\bf 3. The results}

\bigskip

With the use of eq.(3), in this section we provide the results for the $\hat x$ standard deviation, as well as for the  kurtosis for the quantum harmonic Brownian particle.
Two sets of the results are provided:  the exact quantum-mechanical expressions as well as  the decoherence limit, which assumes neglecting the second term in eq.(3). Those results are (separately) compared to the exact classical-physics  results for the harmonic Brownian motion. Along with the variations of the damping coefficient $\gamma$ and the temperature $T$, we also consider the variations due to the oscillator's frequency $\omega$, which is absent from the considerations regarding the free Brownian particle.

\bigskip

{\bf 3.1 The standard deviations}

\bigskip

With the use of eq.(3) it is straightforward but tedious to obtain the standard deviation $\Delta \hat x$, which we overtake from equation (B.2) in Ref. [30] while exchanging the quantities for a rotator with the quantities for the translational motion:

\begin{eqnarray}
&\nonumber&
(\Delta \hat x(t))^2 = {k_BT\over m\omega^2\Omega^2} \left(\Omega^2 +e^{-2\gamma t}
(\omega^2 - \gamma^2 \cosh(2\Omega t) -\gamma\Omega\sinh(2\Omega t))\right)+
\\&&\nonumber
{(\Delta \hat p(0))^2\over m^2\Omega^2} e^{-2\gamma t} \sinh^2(\Omega t) +{(\Delta \hat x(0))^2\over \Omega^2}e^{-2\gamma t}(-\omega^2 \cosh^2(\Omega t)+
\\&&
\gamma^2\cosh(2\Omega t)+\gamma\Omega \sinh(2\Omega t)) + {e^{-2\gamma t}\sigma(0)\over2m\Omega^2}(2\gamma \sinh^2(\Omega t)+
\Omega\sinh(2\Omega t)).
\end{eqnarray}

\noindent In eq.(4): $\Omega^2 = \gamma^2 -\omega^2$,  while the quantum variance $\sigma\equiv\langle \hat  x \hat  p - \hat p \hat  x\rangle - 2\langle \hat  x\rangle\langle \hat p\rangle$.

Placing the classically allowed zero initial moments, $\Delta \hat x(0) = 0, \Delta \hat p(0) = 0$ and $\sigma(0) = 0$, while assuming without any loss of generality, $\langle \hat x(0)\rangle = 0 = \langle \hat p(0)\rangle$, in eq.(4) remains the first term, i.e. the classical expression for $\Delta x(t)$ for a classical Brownian harmonic oscillator [30] (and the references therein):

\begin{equation}
(\Delta  \hat x(t))^2 = {k_BT\over m\omega^2\Omega^2} \left(\Omega^2 +e^{-2\gamma t}
\left(\omega^2 - \gamma^2 \cosh(2\Omega t) -\gamma\Omega\sinh(2\Omega t)\right)\right).
\end{equation}

For the decoherence-limit, neglecting the second term on the rhs of eq.(1), follow the corresponding expression for $\Delta \hat x(t)$. To this end, we directly overtake the equation (C.2) in Ref.[30],
with the quantities for the translational motion:

\begin{eqnarray}
&\nonumber&
(\Delta \hat x(t))^2 = (\Delta \hat x(0))^2 \cos^2{\omega } t
+  {(\Delta \hat p(0))^2 \over m^2\omega^2} \sin^2 {\omega} t +
{\sigma(0) \over 2m\omega} \sin 2{\omega} t
\\&&  + {2\gamma k_BT\over m\omega^2}  t - {\gamma k_BT\over m \omega^3} \sin 2{\omega} t.
\end{eqnarray}

\noindent Of course, being a purely quantum-mechanical effect, the pure decoherence equation (6) does not have a distinguished classical counterpart.

Therefore we compare the exact (equation (4)) and the decoherence-limit (equation (6)) quantum expressions with the (exact) classical equation (5).
Dependence of $(\Delta x(t))^2$ is separately investigated for every parameter, $\gamma, T$ and $\omega$.

Graphical results are presented  in Figure 1 and in Figure 2 (the Log-Log plots), respectively, for the initial values $(\Delta \hat x(0))^2=10^{-7}$, $(\Delta \hat p(0))^2=10^7$ and $\sigma(0)=0.01$, in accordance with the uncertainty relation $[\hat x,\hat p]=\imath\hbar$ (keeping $\hbar = 1$) and the (quantum-mechanical) Cauchy-Schwartz inequality $\sigma\le 2\Delta \hat x \Delta \hat p$,
while a long time interval $t=10$ is chosen. For brevity, we place $\Delta^2 x$ instead of $(\Delta \hat x)^2$.
The choice of the parameter values and ranges as well as of the initial conditions
is made in order to facilitate comparison of the obtained results with the results presented for the free particle in Ref. [22]--as explicitly emphasized in  Figure 1 captions.

\begin{figure*}[!ht]
\centering
    \includegraphics[width=0.3\textwidth]{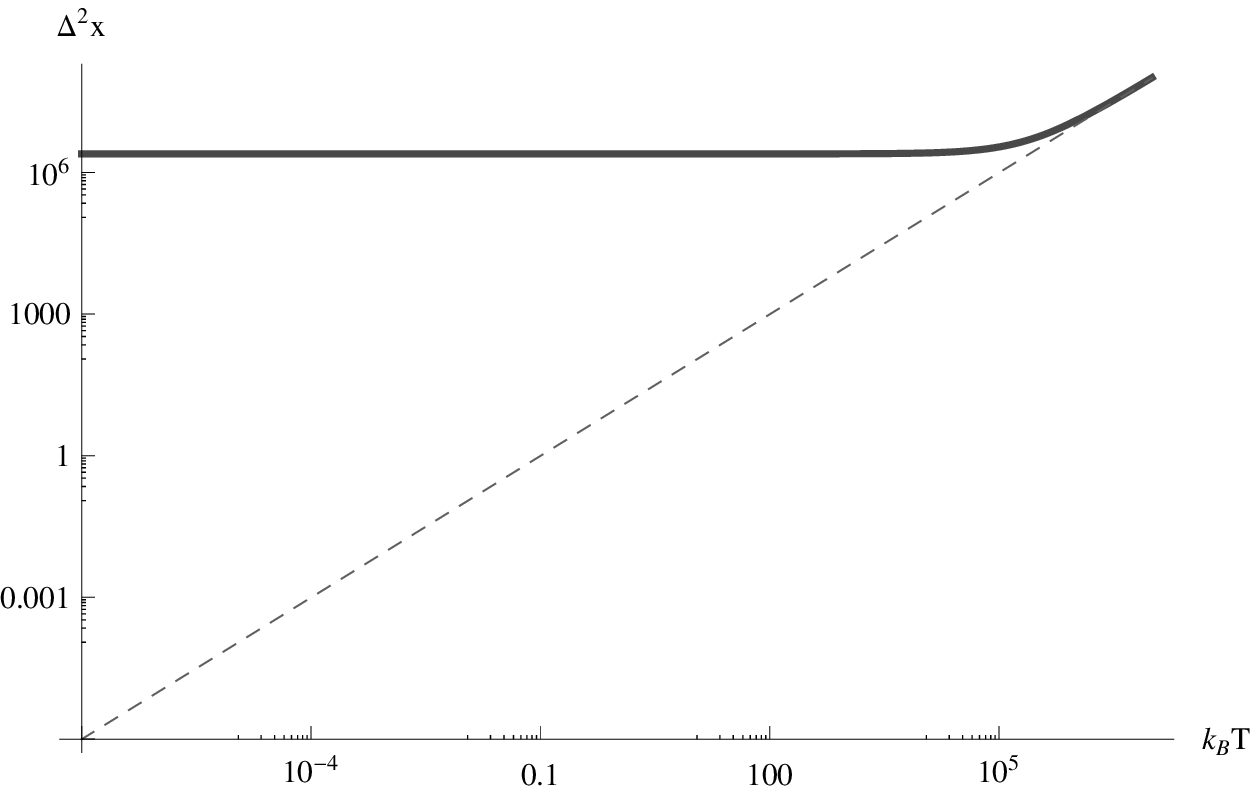}
    \includegraphics[width=0.3\textwidth]{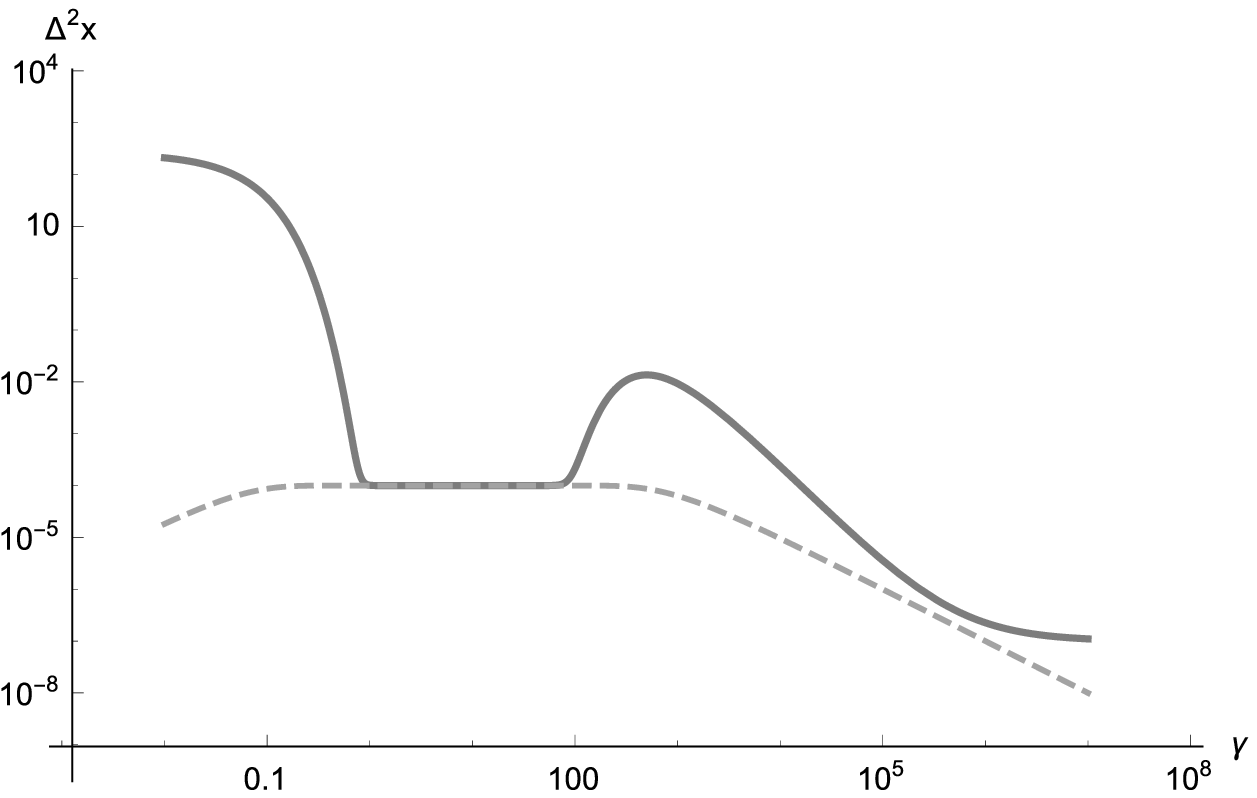}
    \includegraphics[width=0.3\textwidth]{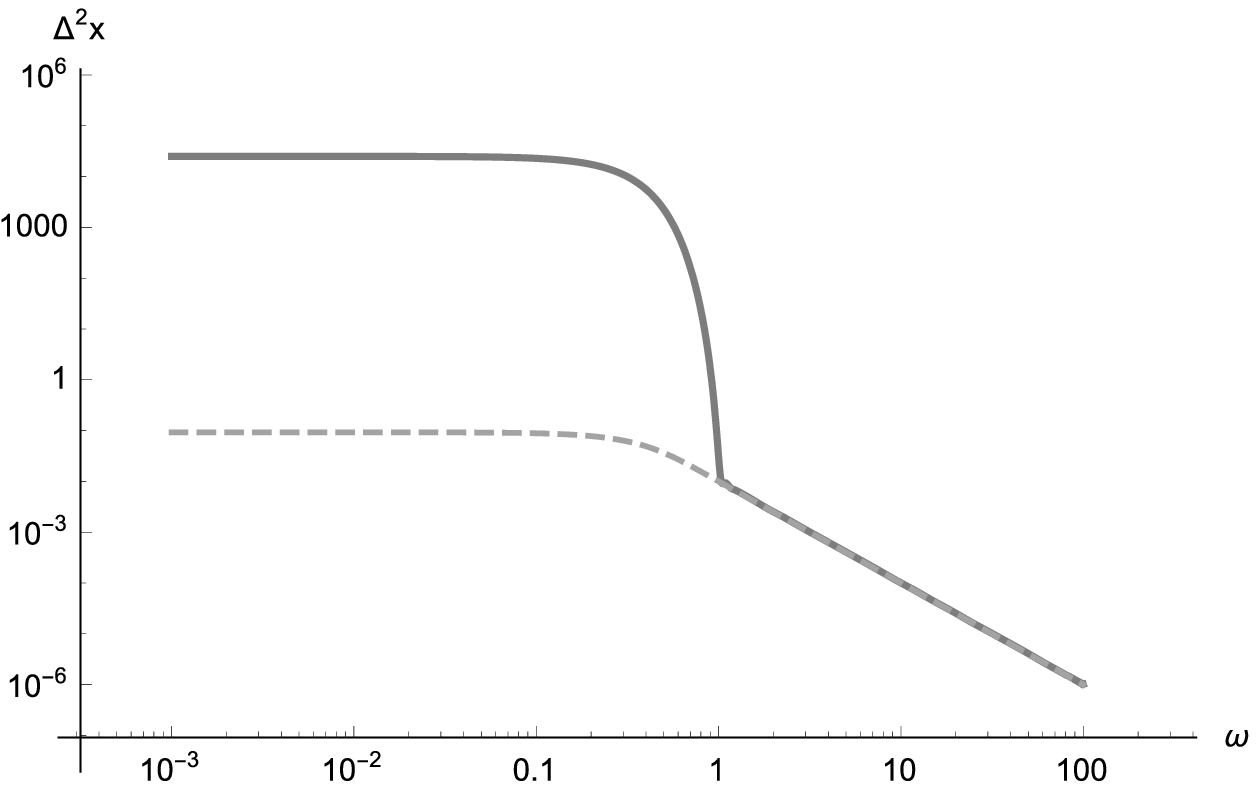}
\caption{Comparison of the exact quantum (solid lines) and the exact classical (dashed lines) expressions. (Left) $\omega=0.1, m=0.1,\gamma=10$, the variable $k_BT\in [10^{-7},10^7]$; (Middle) $\omega=10, m=10, k_BT=0.1$, the variable $\gamma \in [10^{-2},10^7]$; (Right) $m=10, \gamma=1, k_BT=0.1$, the variable $\omega\in [10^{-2}, 10^2]$.}
\end{figure*}

\bigskip

\begin{figure*}[!ht]
\centering
    \includegraphics[width=0.3\textwidth]{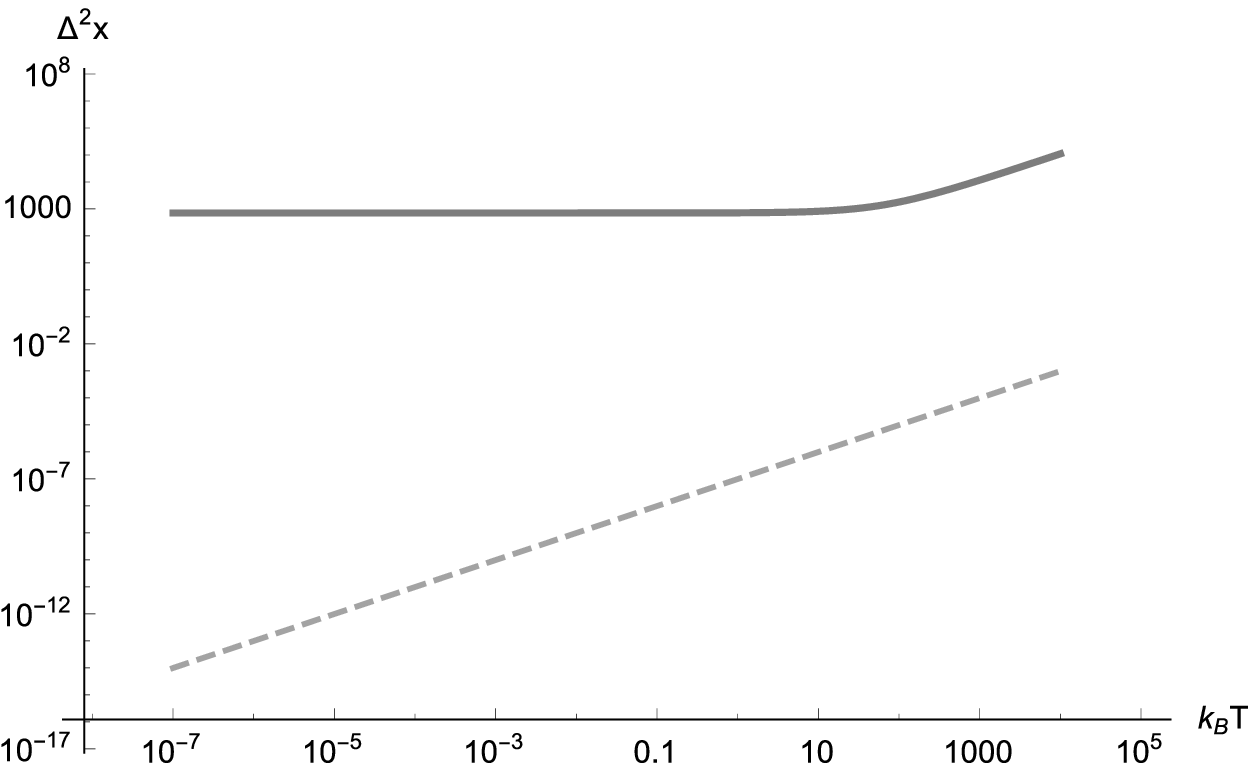}
    \includegraphics[width=0.3\textwidth]{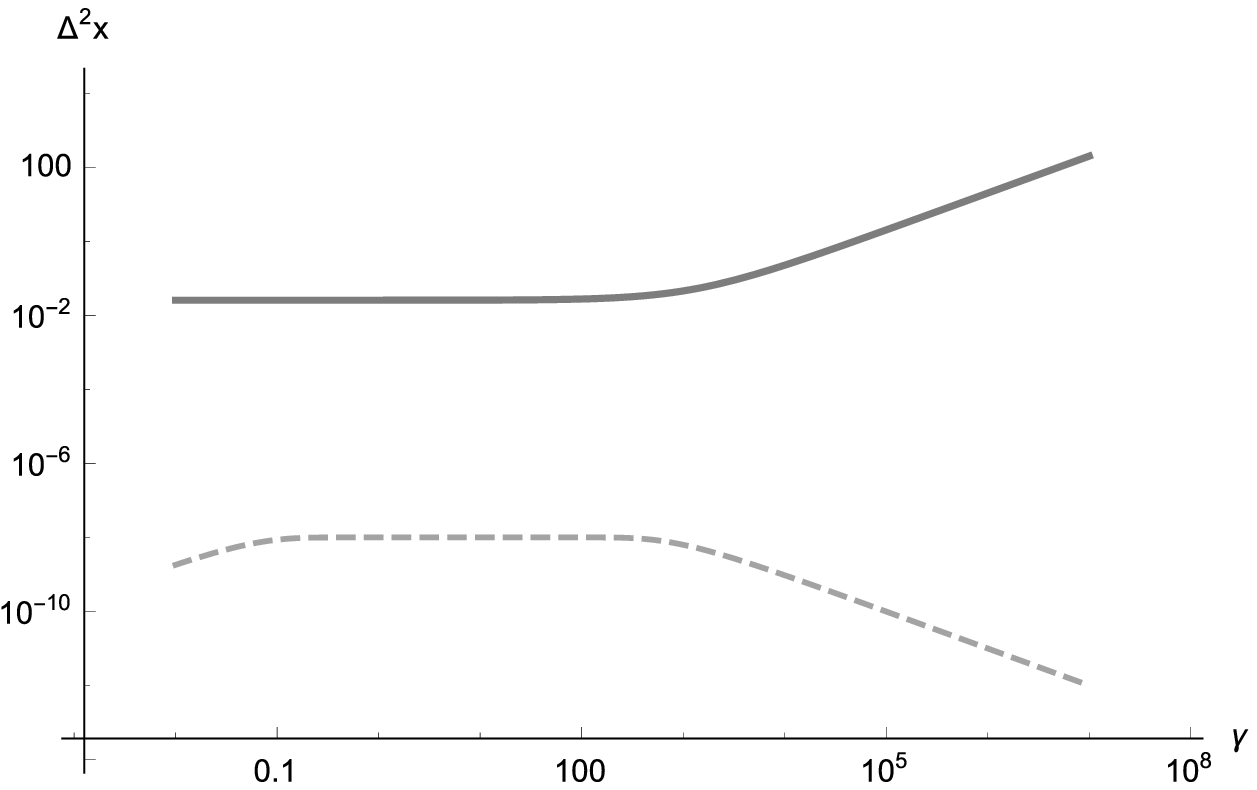}
    \includegraphics[width=0.3\textwidth]{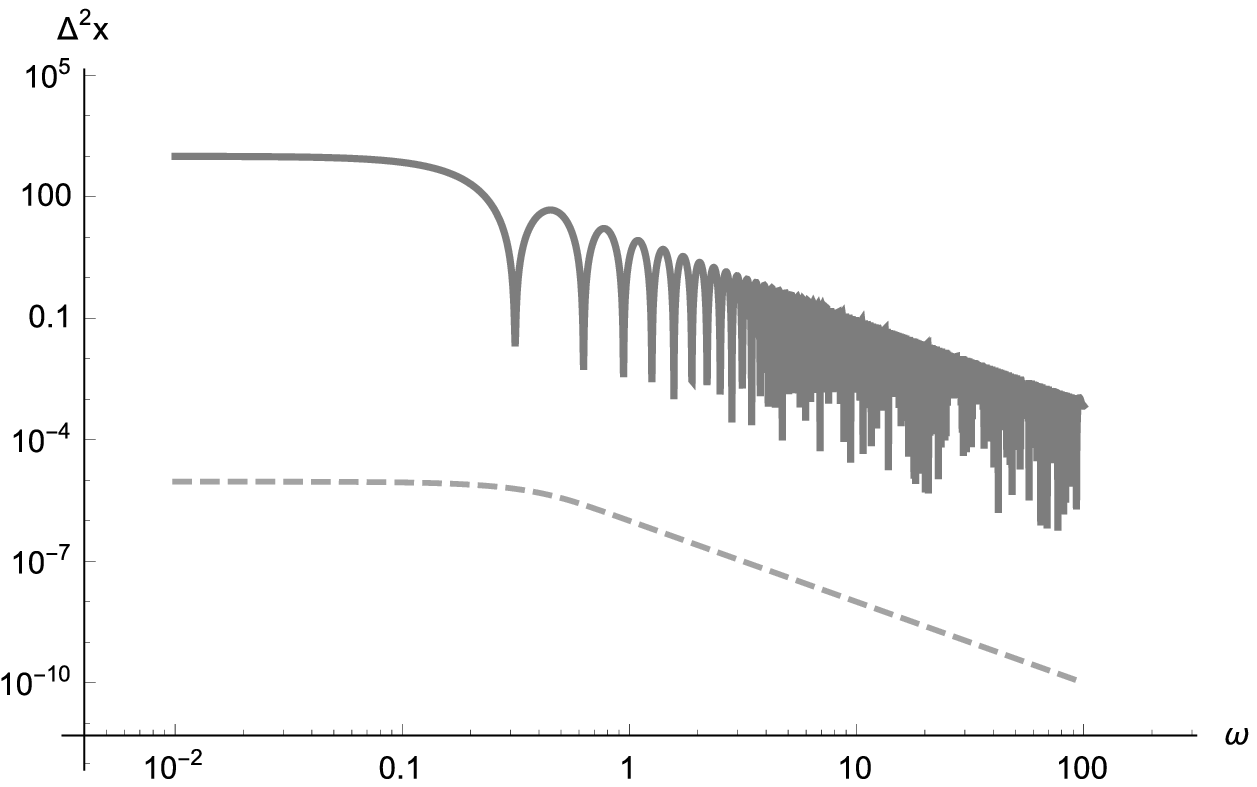}
\caption{Comparison of the quantum decoherence-limit (solid lines) and the exact classical (dashed lines) expressions. The meaning of the plots and of the respective values of the parameters are the same as for Figure 1, except for the choice of the very large mass $m=1000$.}
\end{figure*}

\bigskip

{\bf 3.2 Kurtosis}

\bigskip

Bearing in mind Section 3.1, in this section we only consider the  kurtosis ($\kappa = \langle \hat x^4\rangle / (\langle \hat x^2\rangle)^2$) for the exact quantum-mechanical expression eq.(4).
In Appendix A, we provide the basis for the calculation of the kurtosis for both the free ($\kappa_{free}$) and the harmonic ($\kappa_{harmonic}$) Brownian particle. For both, the free and the harmonic Brownian particle, we obtain (Appendix A) the closed set of the differential equations without a need to use (or calculate) the moments of any higher order. Since we find that the analytical expressions are beyond a succinct  presentation and are physically non-transparent, here we present the results for the chosen values of the parameters as well as of the initial values of the relevant moments.

Figure 3 provides a comparison of the results for $\kappa_{free}$ (dashed line) and $\kappa_{harmonic}$ (solid line) as the functions of time $t$ for the choice of the parameters (underdamped regime): $m=20, \gamma = 0.001, k_BT = 0.38$.
In Chinese stock market there is a price limit rule: the rate of return in a trading day cannot be larger than $\pm 10\%$ comparing with the previous day's closing price, which applies to most stocks in China.
To this end we choose the realistic value of the circular frequency $\omega = 18 \cdot 10^{-3}$min. The time unit for Figure 3 is therefore chosen $1$min.

\begin{figure*}[!ht]
\centering
    \includegraphics[width=0.45\textwidth]{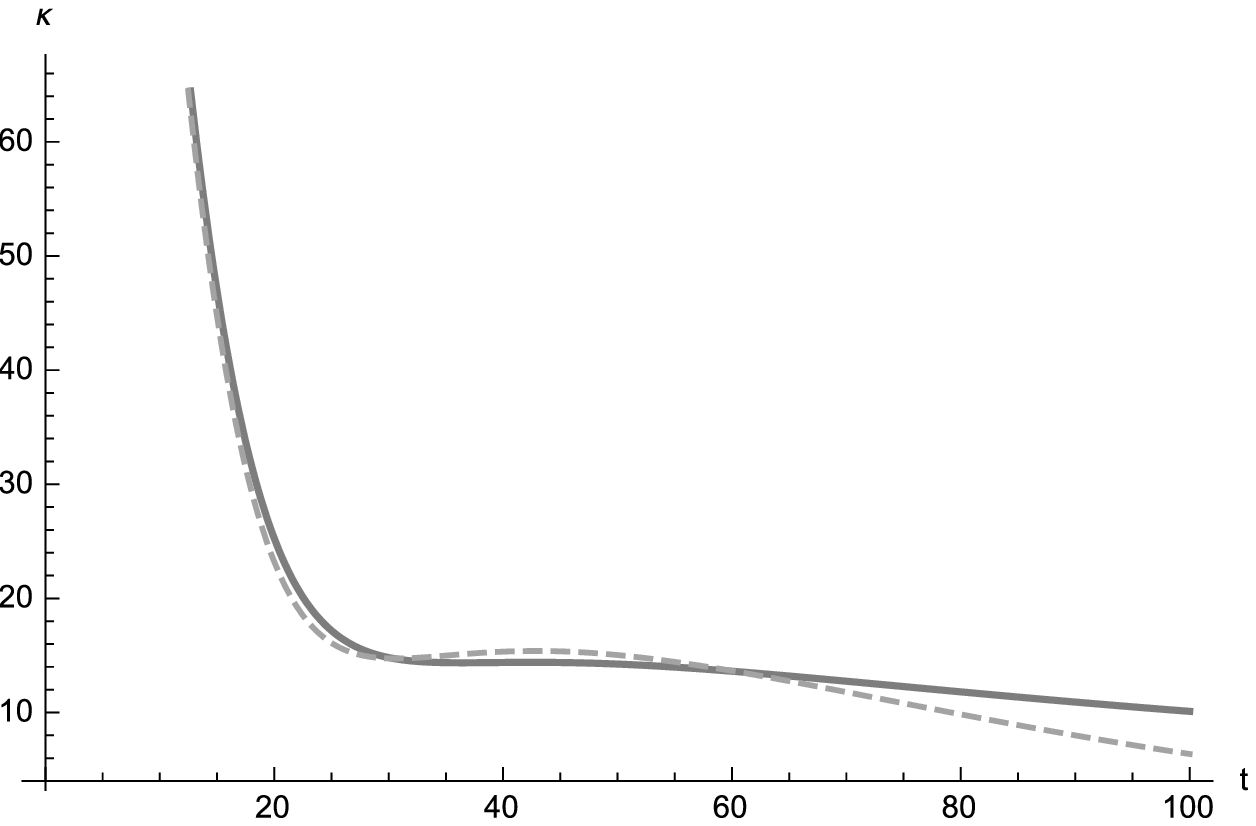}
    \includegraphics[width=0.45\textwidth]{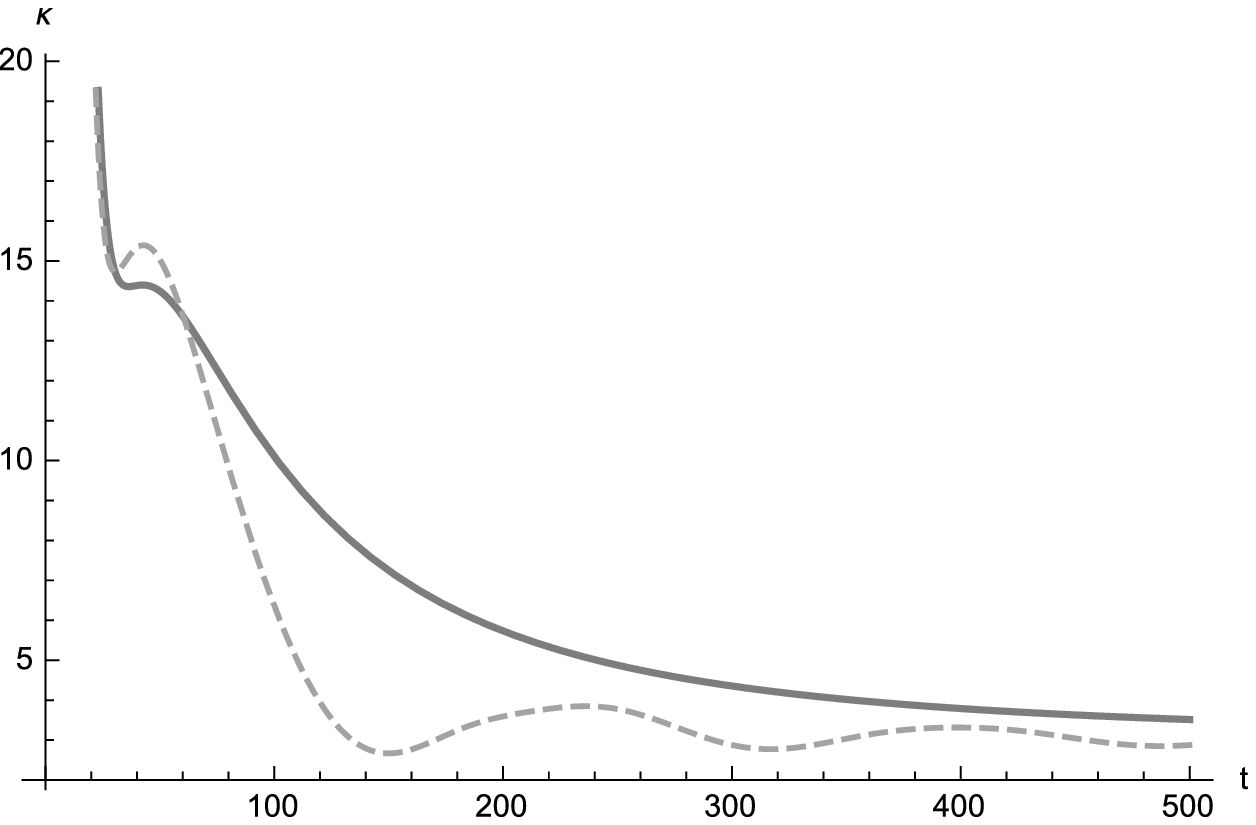}
\caption{Kurtosis for the underdamped regime (assuming $\hbar = 1$) for the two time-intervals. The time scale for both plots is $1$min. The solid line presents the results for the harmonic oscillator while the dotted line presents the results for the free particle. For $t\ge 120$ (Right), the harmonic oscillator exhibits small oscillations around the asymptotic limit of $\kappa = 3$. Approach to the limit for the free particle is  slower and almost monotonic.}
\end{figure*}

The results are rather sensitive to the choice of the initial values for the moments of interest (up to the fourth order) and the presented results are chosen so as to provide the "best" fit of the harmonic-oscillator-model with the evidence data presented by the blue line in Fig.2c in Ref. [22]. The following initial values are used for the first and the second moments:  $\langle \hat x\rangle = 0 = \langle \hat p\rangle$,  $\langle \hat x^2\rangle=1/2=\langle \hat p^2\rangle$ and $\langle \hat x^4\rangle = 50$.

\bigskip

{\bf 3.3 Comments}

\bigskip

Closeness of the quantum and the classical dynamics, Figure 1, exhibits the approxi\-mately-classical dynamics for the oscillator.
Deviation of the quantum from the classical Brownian dynamics is qualitatively a {\it desired behavior} that is
supported by evidence for different stock markets [1-5].

For smaller values of the parameters ($\gamma, T, \omega$), the exact quantum mechanical behaviors, Figure 1, exhibit deviations from the classical counterpart that are much larger than those observed for the
free Brownian motion [22]. While for large $T$ and $\omega$ the behavior is virtually indistinguishable from the classical behavior, for larger values of the $\gamma$ parameter, the deviation is small but observable and quantitatively comparable with the results presented in Figure 2b in Ref. [22].

On the other hand, the approximate quantum behavior presented by Figure 2 is {\it nowhere similar} with the classical counterpart. That is, the large-mass-limit (the decoherence-limit) reveals the behavior that is apart from the classically predicted one; analogous result for the Brownian rigid rotator can be found in Appendix C in Ref. [30]. Inevitably, the quantum decoherence process does {\it not} prove to be sufficient for the classical-like behavior of the quantum Brownian oscillator. As we observe due to the knowledge of the exact classical behavior, only some {\it interplay}  of decoherence and dissipation provides the classically well-known Brownian effect in the context of the quantum mechanical dynamics.  Hence we offer an answer to the question [22] of the relevance of the off-diagonal terms of the density matrix. Bearing in mind the dynamical disappearance of the off-diagonal elements, i.e. $\rho(x,x')\equiv\langle x\vert \hat \rho\vert x'\rangle \propto \exp(-\Gamma(x-x')^2)$, [the only effect in the decoherence-limit], while decoherence cancels out the quantum coherence, this is not sufficient for the classically observed Brownian effect, which {\it requires} also dissipation in the system. In other words, the dynamically disappearing the off-diagonal elements
constitute a necessary but insufficient condition for the classical-like behavior of the quantum harmonic Brownian oscillator.

Figure 3(Right) clearly exhibits a faster approach of $\kappa_{harmonic}$ to the asymptotic Gaussian value of  $\kappa = 3$, which is approximately attained for $t = 150$min.
This approach  is not as smooth as for the free Brownian particle and is qualitatively in accordance with the evidence as presented by the blue line on Fig.2c in Ref.[22].
Comparison of our results with the estimates of the evidence-found results of Fig.2c in Ref. [22] is presented in Table 1.
From Table 1 we can see that, initially, the free Brownian particle gives a better fit with the evidence, while for $t > 60$, the harmonic Brownian particle gives a better fit with the evidence. The values for the evidence-obtained kurtosis are obtained as the (approximate) estimates stemming from Fig.2c in Ref. [22]. Therefore, at least for   the time window $t \in (61, 100)$, we find the better qualitative as well as the quantitative match of the harmonic oscillator model with the evidence than for the case of the free particle.

\begin{table}[h!]
  \begin{center}
    \caption{Comparison of the kurtosis with the evidence-estimated values.}
    \label{tab:table1}
    \begin{tabular}{|l|l|l|l|l|}
      \textbf{} & \textbf{t=40} & \textbf{t=60} & \textbf{t=80} & \textbf{t=100}\\ 
      \hline
      $\kappa_{evidence}$ & 12 & 11 & 7 & 7\\ 
      $\kappa_{free}$ & 14.4 & 13.61 & 11.8 & 10\\ 
      $\kappa_{harmonic}$ & 15.3 & 13.65 & 9.8 & 6.4\\ 
    \end{tabular}
  \end{center}
\end{table}

\bigskip

{\bf 4. Discussion and Conclusion}

\bigskip

We use the Caldeira-Leggett master equation as a ''phenomenological'' equation without resorting to the microscopic details or quantum-mechanical interpretations of the model. This allows us a departure from the standard high-temperature and weak-interaction assumptions [24,25,28].

Observation of a quantum-like effect does not prove or even suggest that the system of interest is of the quantum mechanical nature. As repeatedly emphasized, cf. e.g. [31], quantum-mechanical effects are consequences, i.e. logical implications of (necessary conditions for) the quantum mechanical formalism but not necessarily vice versa. That is, the possible match of the evidence with the quantum mechanical predictions does not imply quantum-mechanical nature of the observed system.

Quantum contributions, Figure 1, increase the volatility measured by the standard deviation. Furthermore, the quantum contributions may even additionally increase due to certain fast actions repeated in short time interval of time. Actually, fast actions which are not accounted for by the CL model, equation (1), typically lead to the increase of the standard deviation. A series of such repeated actions in a short time interval (for which, the quantum corrections must be accounted for) can lead
to a quick, non-negligible increase in $\Delta \hat x$ [30]. This makes the overall dynamics even less predictable.

From Figure 1 we can learn that for large temperature $T$ and for large frequency $\omega$ as well as for not-very-small and not-very-large the damping factor $\gamma$, the dynamics is essentially classical. From figure 2, we can see that for very large mass $m$, the dynamics is never similar to the classical counterpart. Hence one might expect that for sufficiently large $T$ and $\omega$ and for the ''medium'' $\gamma$ and small mass $m$, the quantum corrections can be essentially avoided. However, there is a caveat to this expectation. On the one hand, in the limit of $T\to 0$ and $\gamma = 0$, the system becomes ''closed'', i.e. unitary and therefore deterministic even in the classical limit. On the other hand, in the limit $m\to 0$, equations (4)-(6) give the totally uncontrollable system since in this limit one obtains $\Delta x(t) \to \infty, \forall{t}$. Therefore the ''obvious'' choice of very small $T, \gamma$ and $m$ should be taken with care.

Results presented by Figure 3 are very sensitive to the choice of the higher-momenta initial values, which are used for both the free and the harmonic Brownian particle models. From Figure 3 and from Table 1, we can see the better qualitative (the wavy parts in Figure 3) and  quantitative (Table 1) fit of the harmonic-particle model with the evidence-obtained results than for the case of the free Brownian particle. Therefore we may conclude that the harmonic-oscillator model may be  useful  for description of the realistic stock markets.

The results of Section 3  have the clear economics interpretations.  While  the externally induced fluctuations and damping, quantified by the temperature $T$ and the damping coefficient $\gamma$, respectively, are practically out of control, the capitalization $m$ and the frequency $\omega$ can in principle  be controlled in the realistic situations. From Figure 2 we can learn that too large capitalization drives the system relatively far from the classical model and therefore in opposition with EMH [12].  Figure 1(Right) exhibits that small frequency of the external influence drives the market dynamics further from the classical counterpart. That is,
volatility is considerably larger for smaller frequencies while it practically disappears for the relatively large frequency $\omega$.
Therefore the ''modest'' capitalization and the not-too-frequent external interventions may reduce the volatility and therefore the risk. In order to make this observation more precise, let us
assume that there is an option, or other financial derivative, that can be used to manage the risk,  with the {\it initially small} volatility (standard deviation).
Then a series of such actions {\it performed by a large number of agents in a short time interval}
may lead to a sharp, non-negligible increase of the volatility and
therefore to a sudden break of the initial stability of the market. That is, numerous agents actions performed in a short time window, especially right after the opening of the stock market, inevitably induce increase of the quantum contributions to volatility thus possibly inducing a sharp break of the initial market stability.
This is a classically unknown scenario [30], which reminds us of the essence of the Minsky's financial instability hypothesis [32], that "periods of calm can project a
false sense of security and lure agents into taking a riskier investment, preparing for a crisis" [33]; to this end see also [34].

Collecting the told above, we may say that avoiding the quantum contributions in order to make the stock prices ''more predictable'' may be regarded a kind of the optimization problem, rather than a straightforward procedure with the more-or-less weakly dependent parameters.

Hence, globally, Section 3 clearly emphasizes: there may be additional uncertainty {\it not} predicted by the classical  Brownian model that, while quantitatively approximately fitting to the evidence data (notably Figure 3(Right)), do not offer the simple recipes for avoiding the possible risks. Rather, some optimal strategies are required  for the optimal choice of the parameter values.
Formulation of such strategies is beyond the scope of this paper for at least two reasons. First, we need quantitatively the more elaborated data for comparison. Second, such strategies pose a challenge even for the idealized theoretical models. To this end, the research is ongoing and the results will be presented elsewhere.
Nevertheless, certain lessons are out of question. E.g. by controlling the frequency of the external actions and the capitalization, the market dynamics may partially reveal the effective "temperature" and "damping" on the market
thus providing possibly a deeper insight into the market dynamics and conditions. Finally, very quick and numerous agent actions in a short time window are expected to increase the volatility and therefore make the investments more risky.

We conclude with our expectation, that the progress in quantum-mec\-ha\-nical modeling of real behavior of stock prices may be regarded a kind of justification of the idea of the investor's irrationality as recognized by the behavioral economists. That is, it may be viable to assume that the quantum econophysics studies provide arguments for irrationality of the  agents. Nevertheless, this raises a far-reaching question: if [assumed] irrationality is typical for the economy business, how could it be absent from the other kinds of human endeavors? We believe that this way comes a new broad perspective open by the quantum econophysics studies for different humanistic and social sciences, including sociophysics [36-38]--this time with a more elaborated quantitative criteria.

\bigskip

{\bf Acknowledgements}

\bigskip

This paper is financially supported by Ministry of Science Serbia, grant no 171028.

\bigskip

{\bf Appendix A}

\bigskip

From eq.(3) follows the  set of the differential equations for the moments of the fourth order (to simplify notation, we omit the "hat" operator symbol) that can be presented in the matrix form:

\begin{equation}
{d\over dt} X(t) = \mathcal{M} X(t) + F(t),
\end{equation}

\noindent where the matrix $\mathcal{M}$ reads:

\begin{equation}
\begin{bmatrix}
    0       & 2/m & 0 & 0 & 0  \\
    -2m\omega^2       & -2\gamma & 3/m &  0 & 0 \\
    0 & -2m\omega^2 &  -4\gamma & 2/m &  0 \\
    0 & 0 & -3m\omega^2 & -6\gamma &  2/m   \\
    0 & 0 & 0 & -2m\omega^2 & -8\gamma\\

\end{bmatrix}
\end{equation}

\noindent and the (transposed) vector

\begin{equation}
X^T = (\langle\hat x^4  \rangle, \langle \hat x^3 \hat p + \hat p \hat x^3\rangle, \langle  \hat x^2 \hat p^2 + \hat p^2 \hat x^2 \rangle,  \langle  \hat x \hat p^3 + \hat p^3 \hat x \rangle, \langle \hat p^4\rangle),
\end{equation}

\noindent while the (transposed) nonhomegeneous part reads:

\begin{eqnarray}
&\nonumber&
F^T(t) = (0, 3 \hbar^2/m, -4 \hbar^2 \gamma +
 8 m \gamma k_B T (\Delta \hat x(t))^2,
\\&& -3 \hbar^2 m\omega^2 +
 12 m \gamma k_B T \sigma_{xp}(t),
24 m \gamma k_B T (\Delta \hat p)^2).
\end{eqnarray}

\noindent Appearance of the second moments in eq.(10) makes the set of the differential equations eq.(7) closed--there are no moments of the order larger than four.

The general solution of eq.(7) can be written as

\begin{equation}
X(t) = e^{\mathcal{M} t} X(0) + \int_0^t ds e^{\mathcal{M}(t-s)} F(s).
\end{equation}

\noindent The free Brownian particle is obtained by setting $\omega=0$ in eq.(8) and repeating the same procedure as for the harmonic Brownian particle.

The analytical expressions for $\langle x^4\rangle$ are rather large and non-transparent, for both cases of the free and the harmonic particle. Therefore we only provide the results for the proper choices of the initial values for the moments and for the system parameters ($\gamma, k_BT, \omega$) as described in Section 3.2 of the body text.
The exact analytical expression for the second moment $\langle \hat x^2\rangle$ is well-known for the free Brownian particle, see eq.(3.438) in Ref.[25]:

\begin{eqnarray}
&\nonumber&
(\Delta \hat x(t))^2 = (\Delta \hat x(0))^2 + \left({1- e^{-2\gamma t}\over 2\gamma}\right)^2 {(\Delta \hat p(0))^2\over m^2} + {1 - e^{-2\gamma t} \over 2} \sigma_{x p}(0)+
\\&&
{k_BT\over m\gamma^2} \left(
\gamma t - (1-e^{-2\gamma t})+{1-e^{-4\gamma t} \over 4}
\right).
\end{eqnarray}

\noindent This completes the necessary data for computing the kurtosis for both, the free and the harmonic Brownian particle.

\bigskip

{\bf References}

\bigskip

[1] T. Lux, The stable Paretian hypothesis and the frequency of large returns: an examination of major German stocks, Appl. Fin. Econ. {\bf 6} (1996) 463-475.

[2] P. Gopikrishnan, M. Meyer, L.A. Nunes Amaral, H.E. Stanley, Inverse cubic law for the distribution of stock price variations, Eur. Phys. J. B {\bf 3} (1998) 139-140.

[3] P. Gopikrishnan, V. Plerou, L.A. Nunes Amaral, M. Meyer, H.E. Stanley, Scaling of the distribution of fluctuations of financial market indices, Phys. Rev. E {\bf 60} (1999) 5305-5316.

[4] K. Matia, M. Pal, H. Salunkay and H. E. Stanley, Scale-dependent price fluctuations for the Indian stock market, EPL {\bf 66}  (2004) 909-914.

[5] Z.F. Huang, The first 20 min in the Hong Kong stock market, Physica A {\bf 287} (2000) 405-411.

[6] R. Balvers, Y. Wu, E. Gilliland, Mean Reversion across National Stock Markets and Parametric Contrarian Investment Strategies, J. Finance, {\bf 55} (2000) 745-772.

[7] J.-W. Zhang, Y. Zhang, H. Kleinert, Power tails of index distributions in Chinese stock market, Physica A {\bf 377} (2007) 166-172.

[8] W. Wan, J.-W. Zhang, Long-term memory of the returns in the Chinese stock indices, Front. Phys. China {\bf 3} (2008) 489-494.

[9] P. Gopikrishnan, V. Plerou, X. Gabaix, H.E. Stanley, Phys. Rev. E {\bf 62} (2000) R4493.

[10] V. Plerou, P. Gopikrishnan, L.A. Nunes Amaral, X. Gabaix, H.E. Stanley, Phys. Rev. E {\bf 62} (2000) R3023.

[11] K. Matia, L.A. Nunes Amaral, S.P. Goodwin, H.E. Stanley, Phys. Rev. E {\bf 66} (2002) 045103(R).

[12] E.F. Fama, Random walks in stock market prices, Financ. Anal. J. {\bf 21} (1965) 55-59.

[13] K. Ilinski, Physics of Finance, Wiley, New York, 2001.

[14] B. E. Baaquie, Quantum Finance, Cambridge University Press, Cambridge, 2004.

[15] M. Schaden, Quantum finance, Physica A {\bf 316} (2002) 511-538.

[16] F. Bagarello, Stock markets and quantum dynamics: a second quantized description, Physica A {\bf 386} (2007) 283-302.

[17] F. Bagarello, The Heisenberg picture in the analysis of stock markets and in other sociological contexts, Qual. Quant. {\bf 41} (2007) 533-544.

[18] C. Ye, J. Huang, Non-classical oscillator model for persistent fluctuations in stock markets, Physica A {\bf 387} (2008) 1255-1263.

[19] A. Ataullah, I. Davidson, M. Tippett, A wave function for stock market returns, Physica A, {\bf 388} (2009) 455-461.

[20] C. Zhang, L. Huang, A quantum model for the stock market, Physica A {\bf 389} (2010) 5769-5775.

[21] X. Meng, J.-W. Zhang, J. Xu, H. Guo, Quantum spatial-periodic harmonic model for daily price-limited stock markets, Physica A {\bf 438} (2015) 154-160.

[22] X. Meng, J.-W. Zhang, H. Guo, Quantum Brownian motion model for the stock market, Physica A {\bf 452} (2016) 281-288.

[23] K. Ahn, M. Y. Choi, B. Dai, S. Sohn, B. Yang, Modeling stock return distributions with a quantum harmonic
oscillator, EPL {\bf 120} (2017) 38003.

[24] A.O. Caldeira, A.J. Leggett, Path integral approach to quantum Brownian motion, Physica A {\bf 121} (1983) 587-616.

[25] H.-P. Breuer, F. Petruccione, The Theory of Open Quantum Systems, Clarendon, Oxford, 2002.

[26] K. A. Kim, J. Park, Why do price limits exist in stock markets? A manipulation-based explanation, Europ. Finan. Manage. {\bf 16} (2010) 296.

[27] R. Frisch, Propagation problems and impulse problems in dynamic economics, in: Economic Essays in Honor of Gustav Cassel, George
Allen \& Unwin, London, 1933, pp. 171-205.

[28] L. Ferialdi, Dissipation in the Caldeira-Leggett model, Phys. Rev. A {\bf 95} (2017) 052109.

[29]  E. Joos, H.-D. Zeh, C. Kiefer, D.J.W. Giulini, J. Kupsch, I.-O. Stamatescu, Decoherence and the Appearance of a Classical World in Quantum Theory, Springer-Verlag, Berlin, Heidelberg, 2003.

[30] J. Jekni\' c-Dugi\' c, I. Petrovi\' c,  M. Arsenijevi\' c, M. Dugi\' c, Dynamical stability of the one-dimensional rigid
Brownian rotator: the role of the rotator's spatial size and shape, J. Phys. Cond. Matt. {\bf 40} (2018) 195304.

[31] A. J. Leggett, Macroscopic Quantum Systems and the Quantum Theory of Measurement, Prog. Theor. Phys. Suppl. {\bf 69} (1980) 80-100.

[32] P. H. Minsky, The Financial Instability Hypothesis, L\' evy Economics Institute Working Paper No. 74,
doi:10.2139/ssrn.161024 (1992).

[33] D. Valenti, G. Fazio, B. Spagnolo, Phys. Rev. E  {\bf 97} (2018) 062307.

[34] J. Yellen, Transcript of Chair Yellen’s Press Conference, 18 June 2014, https://www.federalreserve.gov/mediacenter/files
/FOMCpresconf20140618.pdf.

[35] P. Wilmott, Paul Wilmott Introduces Quantitative Finance, John Wiley and Sons, Ltd, 2007.

[36] B. K. Chakrabarti, A. Chakraborti and A. Chatterjee, eds., Econophysics and Sociophysics, WILEY-VCH Verlag GmbH \& Co. KGaA, Weinheim, Germany, 2006.

[37] F. Abergel et al, eds., Econophysics and Sociophysics: Recent Progress and Future Directions, Springer International Publishing AG, Cham, Switzerland, 2017.

[38] R. Kutner et al, Econophysics and sociophysics: Their milestones \& challenges, Physica A {\bf 516} (2019) 240-253.

\end{document}